\documentclass[showpacs,aps,prd,preprint,nofootinbib,showkeys,unsortedaddress,raggedbottom]{revtex4-1}
\pdfoutput=1

\usepackage{bm}
\usepackage{amsmath}
\usepackage{graphicx}
\usepackage[usenames,dvipsnames]{color}
\definecolor{darkblue}{RGB}{0,0,196}
\usepackage[colorlinks=true,linkcolor=darkblue,citecolor=darkblue,urlcolor=darkblue]{hyperref}

\usepackage{slashed}

\def\be {\begin{equation}}
\def\ee {\end{equation}}
\def\bea {\begin{eqnarray}}
\def\eea {\end{eqnarray}}
\def\bc {\begin{center}}
\def\ec {\end{center}}
\def\nn {\nonumber}

\def\sumint{\sum\!\!\!\!\!\!\!\!\!\!\!\!\!\!\!\!\!\int\limits}

\begin{document}

\title{Dilepton rate and quark number susceptibility with the Gribov action}

\author{Aritra Bandyopadhyay}
\affiliation{Theory Division, Saha Institute of Nuclear Physics, 1/AF Bidhan Nagar, Kolkata 700064, India}
\email{aritra.bandyopadhyay@saha.ac.in}

\author{Najmul Haque}
\affiliation{Physics Department, Kent State University, Kent, OH 44242,  United States}
\email{nhaque@kent.edu}

\author{Munshi G Mustafa}
\affiliation{Theory Division, Saha Institute of Nuclear Physics, 1/AF Bidhan Nagar, Kolkata 700064, India}
\email{munshigolam.mustafa@saha.ac.in}

\author{Michael Strickland}
\affiliation{Physics Department, Kent State University, Kent, OH 44242,  United States}
\email{mstrick6@kent.edu}

\begin{abstract}
We use a recently obtained resummed quark propagator at finite temperature which takes into account both the chromoelectric scale $gT$ and the chromomagnetic scale $g^2T$ through the Gribov action.  The electric scale generates two massive modes whereas the magnetic scale produces  a new massless spacelike mode in the medium.
Moreover, the non-perturbative quark propagator is found to contain no discontinuity in contrast to the standard perturbative hard thermal loop approach. Using this non-perturbative quark propagator and self-consistent vertices, we compute the non-perturbative dilepton rate at vanishing three-momentum at one-loop order.  The resulting rate has a rich structure at low energies due to the inclusion of the non-perturbative magnetic scale. We also calculate the quark number susceptibility, which is related to the conserved quark number density fluctuation in the deconfined state. Both the dilepton rate and quark number susceptibility are compared with results from lattice quantum chromodynamics and the standard hard thermal loop approach. Finally, we discuss how the absence of
a discontinuity in the imaginary part of the non-perturbative quark propagator makes the results for both dilepton production and quark number susceptibility dramatically different from those in perturbative approaches and seemingly in conflict with known lattice data.
\end{abstract}

\date{\today} 

\maketitle 

\section{Introduction}
\label{intro}

The ongoing ultra-relativistic heavy-ion collision experiments at RHIC and LHC 
enable us to study the quark-gluon plasma (QGP) which is a deconfined state of 
hadronic matter generated at very high temperatures and/or densities.  Although 
the quark-gluon plasma may be strongly coupled at low temperatures, at high 
temperature there is evidence that resummed perturbation theory can be used to 
understand the properties of the QGP.  To perturbatively study the QGP one needs 
to have an in-depth understanding of the various collective modes.  These 
collective modes can be roughly classified into three types which are associated 
with different thermal scales, namely the energy (or hard) scale $T$, electric 
scale $gT$, and magnetic scale $g^2T$, where $g$ is the strong coupling and $T$ 
is the temperature of the system.  The majority of studies in the literature 
have focused on the hard and electric scales, since the magnetic scale is 
related to the difficult non-perturbative physics of confinement. 

 Based on the Hard-Thermal-Loop (HTL) 
resummations~\cite{braaten1,braaten2,braaten3}, a reorganization of 
finite-temperature perturbation theory called HTL perturbation theory (HTLpt) 
was developed over a decade ago~\cite{andersen1}.  HTLpt deals with the 
intrinsic energy scale $T$ as the hard scale and the electric scale $gT$ as the 
soft scale and has been extensively used to calculate various physical 
quantities associated with the deconfined state of matter.  These quantities 
include: the thermodynamic potential and other relevant quantities associated 
with 
it~\cite{andersen1,andersen7,purnendu1,purnendu2,purnendu3,andersen8,andersen10,
andersen11,najmul1,andersen12,andersen13, 
najmul2,andersen15,najmul4,mogliacci,najmul5,najmul6,najmul7,Strickland:2014zka,
Andersen:2014dua}, photon production rate~\cite{kapusta}, dilepton production 
rate~\cite{yuan,griener}, single quark and quark anti-quark 
potentials~\cite{mustafa1,mustafa3}, photon damping rate~\cite{markus,abada}, 
fermion damping rate~\cite{pisarski,peigne}, gluon damping 
rate~\cite{pisarski1,pisarski2}, plasma 
instabilities~\cite{mrowc,strickland1,strickland2}, jet energy 
loss~\cite{markus1,markus2,markus3,strickland7,strickland8,mustafa5}, lepton  
asymmetry during 
 leptogenesis~\cite{kiessig1,kiessig2}, and thermal axion 
production~\cite{graf}. 

Although HTLpt seems to work well at a temperature of approximately 2 $T_c$ and 
above, where $T_c \sim 160$ MeV is the pseudo-critical temperature for the QGP 
phase transition, the time-averaged temperature of the QGP generated at RHIC and 
LHC energies is quite close to $T_c$.  Near $T_c$, the running coupling $g$ is 
large and the QGP could therefore be completely non-perturbative in this 
vicinity of the phase diagram.  In order to make some progress at these 
temperatures, it is necessary to consider the non-perturbative physics 
associated with the QCD magnetic scale in order to assess its role.  
Unfortunately, the magnetic scale is still a challenge for the theoreticians to 
treat in a systematic manner since, although its inclusion eliminates infrared 
divergences, the physics associated with the magnetic scale remains completely 
non-perturbative~\cite{nadkarni}.  The fact that the $\mathcal{O}(g^2T)$ 
correction to the Debye mass receives non-perturbative contributions indicates 
that the background physics is fundamentally 
non-perturbative~\cite{arnold_yaffe}.  The physics in the magnetic sector is 
described by a dimensionally reduced three-dimensional Yang-Mills theory and the 
non-perturbative nature of the physics in this sector is related with the 
confining properties of the theory.  

Lattice QCD (LQCD) provides a first principles based method that can take into 
account the non-perturbative effects of QCD.  Lattice QCD has been used to probe 
the behavior of QCD in the vicinity of $T_c$, where matter undergoes a phase 
transition from the hadronic phase to the deconfined QGP phase. At this point, 
the QCD thermodynamic functions and some other relevant quantities associated 
with the fluctuations of conserved charges at finite temperature and zero 
chemical potential have been very reliably computed using LQCD (see  
e.g.~\cite{borsanyi2,borsanyi3,bnlb1,bnlb2,milc,hotqcd1,hotqcd2,tifr}).  In 
addition, quenched LQCD has also been used to study the structure of vector 
meson correlation functions.  Such studies have provided critically needed 
information about the thermal dilepton rate and various transport coefficients 
at zero momentum~\cite{ding,kaczmarek1,arts2,laermann} and finite 
momentum~\cite{arts5}.  

Calculations in LQCD proceed by evaluating the Euclidean time correlation 
function only for a discrete and finite set of Euclidean times. To obtain the 
dilepton rate, one needs to perform an analytic continuation of the correlator 
from discrete Euclidean times to reconstruct the vector spectral function in 
continuous real time. However, this is an ill-posed problem.   To proceed, the 
spectral function and hence the dilepton rate in continuous real time can be 
obtained  from the correlator in discrete Euclidean times through a 
probabilistic interpretation based on the maximum entropy method 
(MEM)~\cite{mem1,mem2,mem3}, which requires an ansatz for the spectral 
function.  Employing a free-field spectral function as an ansatz, the spectral 
function in the quenched approximation of QCD was obtained earlier and found to 
approach zero in the low-energy limit~\cite{laermann}.  In the same work, the 
authors found that the lattice dilepton rate approached zero at low invariant 
masses~\cite{laermann}.   In a more recent LQCD calculation with larger lattice 
size, the authors used a Breit-Wigner (BW) form for low-energies plus a 
free-field form for high-energies as their ansatz for the spectral 
function~\cite{ding}.  The low-energy BW form of their ansatz gave a finite 
low-energy spectral function and low-mass dilepton rate.  This indicates that 
the computation of low-mass dilepton rate in LQCD is indeed a difficult task and 
is also not very clear if there are structures in the low-mass dilepton rate 
similar to those found in the HTLpt calculation~\cite{yuan}.

Given the uncertainty associated with lattice computation of dynamical 
quantities, e.g. spectral functions,  dilepton rate, and transport coefficients, 
it is desirable to have an alternative approach to include non-perturbative 
effects that can be handled in a similar way as in resummed perturbation theory. 
A few such approaches are available in the literature:  one approach is a 
semi-empirical way to incorporate non-perturbative aspects by introducing a 
gluon condensate\footnote{An important aspect of the phase structure of QCD is 
to understand the effects of different condensates, which serve as order 
parameters of the broken symmetry phase.  These condensates are non-perturbative 
in nature and their connection with bulk properties of QCD matter is provided by 
LQCD.  The gluon condensate has a potentially substantial impact on the bulk 
properties, e.g., on the equation of state of QCD matter, compared to the quark 
condensate.} in combination with the Green functions in momentum space, which 
has been proposed in e.g. Refs.~\cite{schaefer1,schaefer2,peshier,pc1,pc2,pc3}. 
In this approach, the effective $n$-point functions are related by 
Slavnov-Taylor (ST) identities which contain gluon condensates in the deconfined 
phase as hinted from lattice measurements in pure-glue QCD~\cite{boyd}.  The 
dispersion relations with dimension-four gluon condensates in medium exhibits 
two massive modes~\cite{schaefer1} (a normal quark mode and a plasmino mode) 
similar to HTL quark dispersion relations.  This feature leads to sharp 
structures (van Hove singularities, energy gap, etc.) in the dilepton production 
rates~\cite{schaefer3,peshier} at zero momentum, qualitatively similar to the 
HTLpt rate~\cite{yuan}. 

Using quenched LQCD, Refs.~\cite{kitazawa1,kitazawa2} calculated the 
Landau-gauge quark propagator and its corresponding spectral function by 
employing a two pole ansatz corresponding to a normal quark and a plasmino mode
following the HTL dispersion relations~\cite{yuan}. In a very recent 
approach~\cite{kim}, a Schwinger-Dyson (SD) equation  
has been constructed with the aforementioned Landau-gauge propagator obtained 
using quenched LQCD~\cite{kitazawa1,kitazawa2} and a vertex function related 
through ST identity.  Using this setup the authors computed the dilepton rate 
from the deconfined phase and found that it has the characteristic van-Hove 
singularities 
but does not have an energy gap. 

In a very recent approach~\cite{nansu1} quark propagation in a deconfined medium 
including both electric- and magnetic-mass effects has also been studied by 
taking into account the non-perturbative magnetic screening scale by using the 
Gribov-Zwanziger (GZ) action~\cite{Gribov1,zwanziger1}, which regulates the 
magnetic IR behavior of QCD. Since the gluon propagator with the GZ action is IR 
regulated, this mimics confinement, making the calculations more compatible with 
results of LQCD and functional methods~\cite{mass}.  Interestingly, the 
resulting HTL-GZ quark collective modes consist of two massive modes (a normal 
quark mode and a plasmino mode) similar to the standard HTL dispersions along 
with a \textit{new} massless spacelike excitation which is directly related to 
the incorporation of the magnetic scale through the GZ action. This new quark 
collective excitation results in a long range correlation in the system, which 
may have important consequences for various physical quantities relevant for the 
study of deconfined QCD matter. In light of this, we would like to compute the 
dilepton production rate and the quark number susceptibility (QNS) associated 
with the conserved number fluctuation from the deconfined QGP using the 
non-perturbative GZ action.  

This paper is organized as follows. In sec.~\ref{setup} we briefly outline the setup for quark propagation 
in a deconfined medium using GZ action. In sec.~\ref{dilepton} we calculate the non-perturbative dilepton rate and 
discuss the results.  Sec.~\ref{qns} describes the computation and results of non-perturbative QNS. In sec. 5 we 
summarize and conclude.

\section{Setup}
\label{setup}

We know that gluons play an important role in confinement. In the GZ 
action~\cite{Gribov1,zwanziger1} the issue of confinement is usually tackled 
kinematically with the gluon propagator in covariant gauge taking the 
form~\cite{Gribov1,zwanziger1}
\bea
D^{\mu\nu}(P)=\left[\delta^{\mu\nu}-(1-\xi)\frac{P^\mu P^\nu}{P^2}\right]\frac{P^2}{P^4+\gamma_G^4}\, ,
\label{modified_gluon_prop}
\eea
where the four-momenta $P = (p_0, \vec p)$,  $\xi$ is the gauge parameter, and 
$\gamma_G$ is called the {\em Gribov parameter}. Inclusion of the term involving 
$\gamma_G$ in the denominator moves the poles of the gluon propagator off the 
energy axis so that there are no asymptotic gluon modes.  Naturally, to maintain 
the consistency of the theory, these unphysical poles should not be considered 
in the exact correlation functions of gauge-invariant quantities.  This suggests 
that the gluons are not physical excitations.  In practice, this means that the 
inclusion of the Gribov parameter results in the effective confinement of 
gluons.  

In QCD, the Gribov ambiguity typically results in multiple gauge-equivalent 
copies and, as a result, it renders perturbative QCD calculations ambiguous.  
However, the dimensionful Gribov parameter appearing above can acquire a 
well-defined meaning if the topological structure of the $SU(3)$ gauge group is 
made to be consistent with the theory.  Very recently, this has been argued and 
demonstrated by Kharzeev and Levin~\cite{kharzeev1} by taking into account the 
periodicity of the $\theta$-vacuum~\cite{jackiw} of the theory due to the 
compactness of the $SU(3)$ gauge group.  The recent work of Kharzeev and Levin 
indicates that the Gribov term can be physically interpreted as the topological 
susceptibility of pure Yang-Mills theory and that confinement is built into the 
gluon propagator in Eq.~(\ref{modified_gluon_prop}), indicating non-propagation 
of color charges at long distances and screening of color charges at long 
distances in the running coupling.  This also reconciles the original view 
Zwanziger had regarding $\gamma_G$ being a statistical parameter 
\cite{zwanziger1}.  In practice, $\gamma_G$ can be self-consistently determined 
using a one-loop gap equation and at asymptotically high temperatures it takes 
the following form~\cite{nansu1,zwanziger2,nansu2} 
\bea
\gamma_G = \frac{D-1}{D}\frac{N_c}{4\sqrt{2}\pi}g^2T,  \label{Gribov_para}
\eea
where $D$ is the dimension of the theory and $N_c$ is the number of 
colors.\footnote{Equation (\ref{Gribov_para}) is a one-loop result.  In the 
vacuum, the two-loop result has been determined \cite{Gracey:2005cx} and the 
Gribov propagator form (\ref{modified_gluon_prop}) is unmodified.  Only 
$\gamma_G$ itself is modified to take into account the two-loop correction.  To 
the best of our knowledge, this would hold also at finite temperature.} The 
one-loop running strong coupling, 
$g^2=4\pi\alpha_s$, is 
\bea
g^2(T)=\frac{48\pi^2}{(33-2N_f)\ln\left (\frac{Q^2_0}{\Lambda_0^2} \right )} , \label{alpha_s}
\eea
where $N_f$ is the number of quark flavors and $Q_0$ is the renormalization 
scale, which is usually chosen to be $2\pi T$ unless specified.  We fix the 
scale $\Lambda_0$ by requiring that $\alpha_s$(1.5 GeV) = 0.326, as obtained 
from lattice measurements~\cite{alphas_lat}. For one-loop running, this 
procedure gives $\Lambda_0 = 176$ MeV. 

To study the properties of a hot QGP using (semi-)perturbative methods, the 
effective quark propagator is an essential ingredient.  After resummation, the 
quark propagator can be expressed as 
\bea
i S^{-1}(P) &=& \slashed{P}-\Sigma(P), \label{resum_prop} 
\eea
where $\Sigma(P)$ is the quark self energy. One can calculate $\Sigma$ using the 
modified gluon propagator 
(\ref{modified_gluon_prop}) in the high-temperature limit to 
obtain~\cite{nansu1}
\bea
\Sigma(P) &=& (ig)^2C_F \sumint_{\;\;\;\;\;\;\;\;\;\;\;\;\{K\}} \!\! \gamma_\mu S_f(K) \gamma_\nu D^{\mu\nu}(P-K)
\approx -(ig)^2C_F\sum_\pm\int\limits_0^\infty \frac{dk}{2\pi^2}k^2\int \frac{d\Omega}{4\pi} \nn \\
&& \times \frac{\tilde{n}_\pm(k,\gamma_G)}{4E_\pm^0}\left[\frac{i\gamma_0+\hat{\bf{k}}\cdot\bf{\gamma}}{iP_0+k-E_\pm^0+
\frac{\bf{p}\cdot\bf{k}}{E_\pm^0}}
 +\frac{i\gamma_0-\hat{\bf{k}}\cdot\bf{\gamma}}{iP_0-k+E_\pm^0-\frac{\bf{p}\cdot\bf{k}}{E_\pm^0}}\right],
\label{modified_quark_self}
\eea
where $\Sigma_{\{K\}}\!\!\!\!\!\!\!\!\!\!\!\!\!\int\;\;\;\;\;\;$ is a fermionic sum-integral, $S_f(K)$ is the bare quark propagator, and 
\bea
\tilde{n}_\pm(k,\gamma_G)&\equiv& n_B\!\left(\sqrt{k^2 \pm i\gamma_G^2}\right)+n_F(k) \nn ,\\
E_\pm^0 &=& \sqrt{k^2 \pm i\gamma_G^2}\ , \label{freq}
\eea
where $n_B$ and $n_F$ are Bose-Einstein and Fermi-Dirac distribution functions, respectively.
The  modified  thermal quark  mass in presence  of the  Gribov term can also be obtained  as
\bea
m_q^2(\gamma_G) = \frac{g^2C_F}{4\pi^2}\sum_\pm\int\limits_0^\infty dk \, \frac{k^2}{E_\pm^0} \, \tilde{n}_\pm(k,\gamma_G).
\label{tmass}
\eea
Using the modified quark self energy given in Eq.~(\ref{modified_quark_self}), it is now easy to write down 
the modified effective quark propagator in presence of the Gribov term as 
\bea
i S^{-1}(P) &=& A_0\gamma_0 - A_s \gamma\cdot \hat{\bf{p}}, \label{propa0as}
\eea
where, keeping the structure typically used within the HTL approximation, $A_0$ and $A_s$ are defined as~\cite{nansu1}
\bea
A_0(\omega,p) &=& \omega-\frac{2g^2C_F}{(2\pi)^2}\sum_\pm\int dk \, k \, \tilde{n}_\pm(k,\gamma_G) 
\left[Q_0(\tilde{\omega}_1^\pm,p)+Q_0(\tilde{\omega}_2^\pm,p)\right]\nn , \\
A_s(\omega,p) &=& p+\frac{2g^2C_F}{(2\pi)^2}\sum_\pm\int dk \, k \, \tilde{n}_\pm(k,\gamma_G) 
\left[Q_1(\tilde{\omega}_1^\pm,p)+Q_1(\tilde{\omega}_2^\pm,p)\right] . \label{a0as}
\eea
Here the modified frequencies are defined as $\tilde{\omega}_1^\pm \equiv E_\pm^0(\omega+k-E_\pm^0)/k$ and 
$\tilde{\omega}_2^\pm \equiv E_\pm^0(\omega-k+E_\pm^0)/k$. The Legendre functions of 
the second kind, $Q_0$ and $Q_1$, are
\bea
Q_0(\omega,p) &\equiv& \frac{1}{2p}\ln \frac{\omega+p}{\omega-p} \\
Q_1(\omega,p) &\equiv& \frac{1}{p}(1-\omega Q_0(\omega,p)). \label{legd}
\eea
Using the {helicity representation, the modified effective fermion propagator can also be written as 
\bea
i S(P) &=& \frac{1}{2} \frac{(\gamma_0 - \gamma\cdot \hat{\bf{p}})}{D_+}
+ \frac{1}{2} \frac{(\gamma_0 + \gamma\cdot \hat{\bf{p}})}{D_-} , \label{hprop}
\eea
where $D_\pm$ are obtained as
\bea
D_+(\omega,p,\gamma_G) &=& A_0(\omega,p) - A_s(\omega,p)\nn 
= \omega - p - \frac{2g^2C_F}{(2\pi)^2}\sum_\pm\int dk k \tilde{n}_\pm(k,\gamma_G) \nonumber \\
&& \hspace{2cm} \times 
\left[Q_0(\tilde{\omega}_1^\pm,p)+ Q_1(\tilde{\omega}_1^\pm,p)
+Q_0(\tilde{\omega}_2^\pm,p)+ Q_1(\tilde{\omega}_2^\pm,p)\right] ,\nn\\
D_-(\omega,p,\gamma_G) &=& A_0(\omega,p) + A_s(\omega,p)\nn 
= \omega + p - \frac{2g^2C_F}{(2\pi)^2}\sum_\pm\int dk k \tilde{n}_\pm(k,\gamma_G) \nonumber \\
&& \hspace{2cm} \times 
\left[Q_0(\tilde{\omega}_1^\pm,p)- Q_1(\tilde{\omega}_1^\pm,p)
+Q_0(\tilde{\omega}_2^\pm,p)- Q_1(\tilde{\omega}_2^\pm,p)\right]. \label{dpm}
\eea

\begin{figure}[t]
\begin{center}
\includegraphics[width=0.32\linewidth]{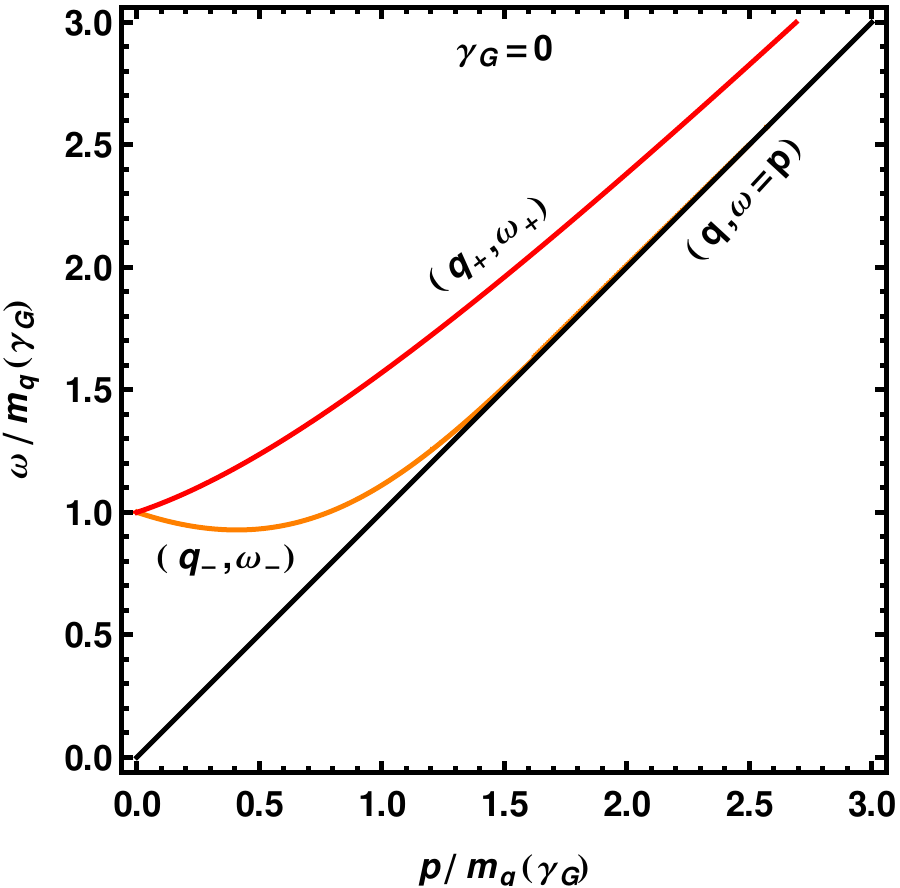}
\includegraphics[width=0.32\linewidth]{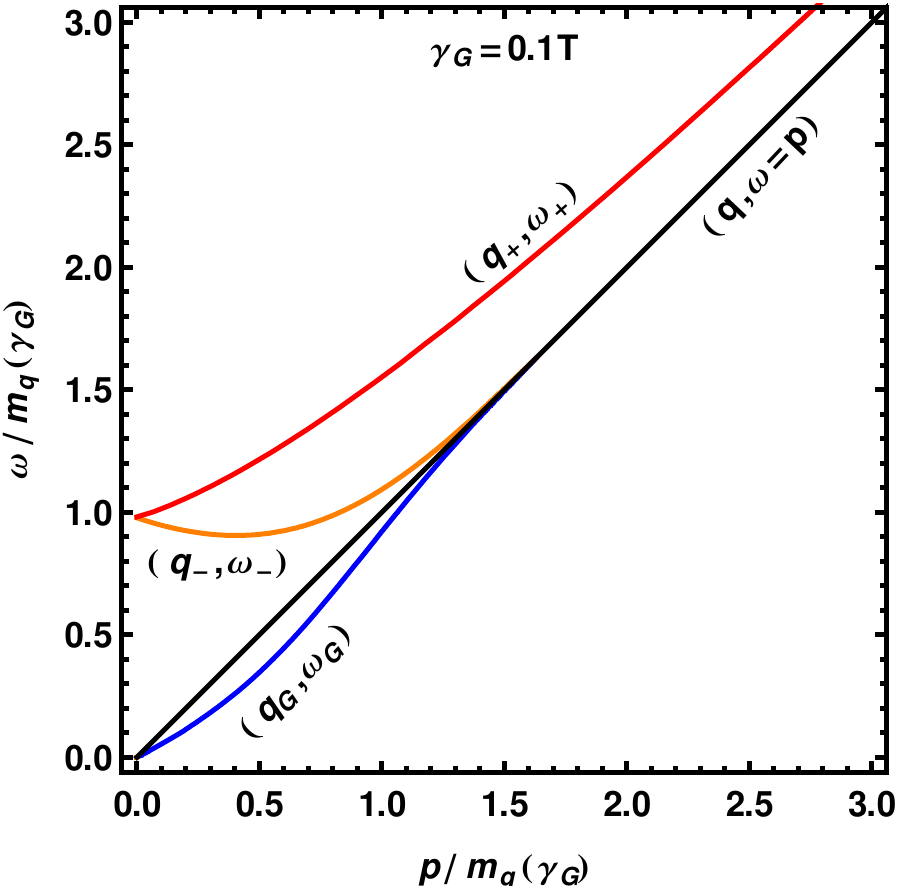}
\includegraphics[width=0.32\linewidth]{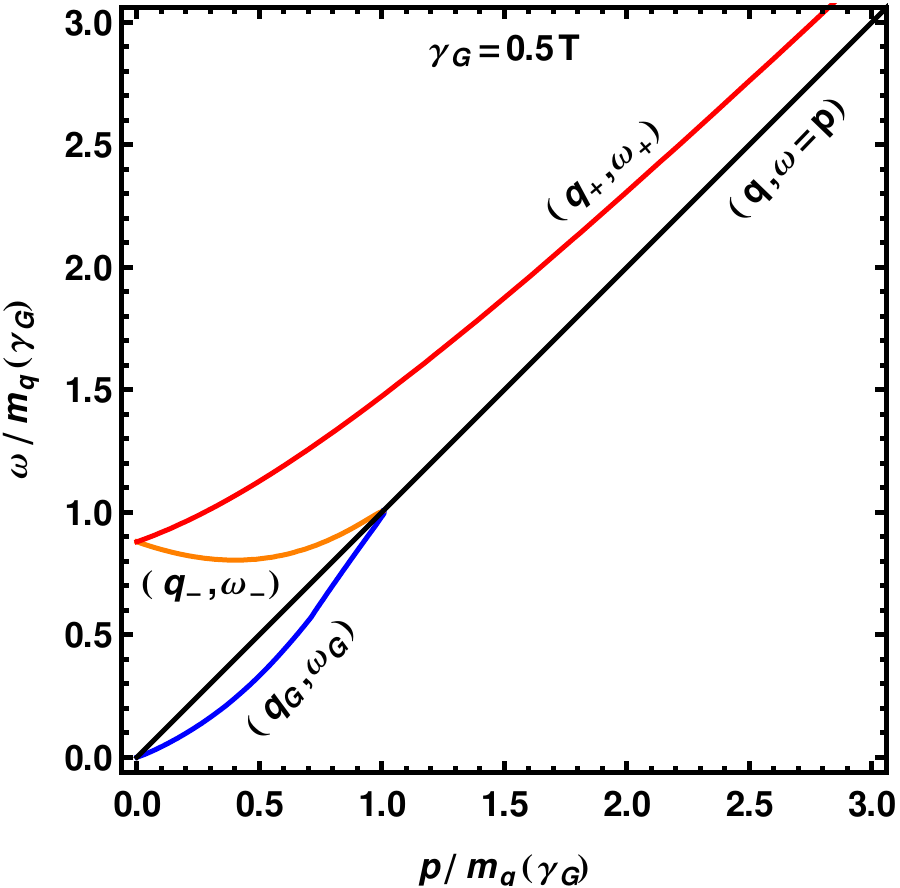} 
\end{center}
\caption{Plot of the dispersion relations for different values of $\gamma_G$. In 
the parenthesis, the first one represents a collective excitation mode whereas 
the second one is the corresponding energy of that mode.}
\label{disp_rel}
\end{figure}

Solving for the zeros of $D_{\pm}^{-1}(P,\gamma_G)$ gives the dispersion 
relations for the collective excitations in the medium.  In Fig.~\ref{disp_rel} 
we show the resulting dispersion relations for three different values of the 
Gribov parameter $\gamma_G$.  In absence of the Gribov term (\textit{i.e,} 
$\gamma_G=0$), there are two massive modes corresponding to a normal quark mode 
$q_+$ with energy $\omega_+$ and a long wavelength plasmino mode $q_-$ with 
energy $\omega_-$ that quickly approaches free massless propagation in the 
high-momentum limit. These two modes are similar to those found in the HTL 
approximation~\cite{yuan}.  With the inclusion of the Gribov term, there is a 
massless mode $q_G$ with energy $\omega_G$, in addition to the two massive 
modes, $q_+$ and $q_-$~\cite{nansu1}. The extra mode $q_G$ is due to the 
presence of the magnetic screening scale.  
This new massless mode is lightlike at large momenta.\footnote{The slope of the 
dispersion relation for this 
extra  massless spacelike mode $q_G$ exceeds unity in some domain of momentum. 
Thus, the group velocity, $d\omega_G/dp$, is superluminal for the spacelike 
mode $q_G$ and approaches the light cone ($d\omega/dp=1$)  from above at high momentum.  
Since the mode is spacelike, there is no causality problem.  Instead, this represents 
anomalous dispersion in the presence of GZ action which converts Landau damping 
into amplification of the spacelike dispersive mode.}  In this context, we note that in Ref.~\cite{pc3}, 
such an extra massive mode with significant spectral width was observed near 
$T_c$ in presence of dimension-four gluon condensates~\cite{pc3} in addition to 
the usual propagating quark and plasmino modes.  The existence of this extra 
mode could affect lattice extractions of the dilepton rate since even the most 
recent LQCD results~\cite{kitazawa1,kitazawa2} assumed that there were only two 
poles (a quark mode and a plasmino mode) inspired by the HTL approximation.

In  HTL approximation ($\gamma_G=0$) the propagator contains a discontinuity in 
complex plane stemming from the logarithmic terms in (\ref{dpm}) due to 
spacelike momentum $\omega^2<p^2$.  Apart from two collective excitations 
originating from the in-medium dispersion as discussed above, there is also a 
Landau cut contribution in the spectral representation of the propagator due to 
the discontinuity in spacelike momentum.  On the other hand, for $\gamma_G \neq 
0$ the individual terms in (\ref{dpm}) possess  discontinuities at spacelike 
momentum but canceled out when all terms are summed owing the fact that the 
poles come in complex-conjugate pairs.  As a result, there is no discontinuity 
in the complex plane.\footnote{Starting from the Euclidean expression 
(\ref{modified_quark_self}), we have numerically checked for discontinuities and 
found none.  We found some cusp-like structures at complex momenta, but $\Sigma$ 
was found to be $C^0$-continuous everywhere in the complex plane.}   This 
results in disappearance of the Landau cut contribution in the spectral 
representation of the propagator in spacelike domain. It appears as if the 
Landau cut contribution in spacelike domain  for $\gamma_G=0$ is replaced by 
massless spacelike dispersive mode in presence of  magnetic scale ($\gamma_G\ne 
0$). So the spectral function corresponding to the propagator $D_{\pm}^{-1}$  
for $\gamma_G \neq 0$ has only pole contributions.  As a result, one has

\bea
\rho_\pm^G(\omega,p)= \frac{\omega^2-p^2}{2m_q^2(\gamma_G)}\left[\delta(\omega\mp\omega_+)+\delta(\omega\pm\omega_-)
+\delta(\omega\pm\omega_G)\right], \label{gspect}
\eea
where  $D_+$ has poles at $\omega_+$, $-\omega_-$, and $-\omega_G$ 
and $D_-$ has poles at $\omega_-$, $-\omega_+$, and $\omega_G$ with a prefactor,  $(\omega^2-p^2)/2m_q^2(\gamma_G)$,  as the residue.  



At this point we would like to mention that the non-perturbative quark spectral 
function obtained using the quark propagator analyzed in the quenched LQCD 
calculations of Refs.~\cite{kitazawa1,kitazawa2,kim} and utilizing gluon 
condensates in Refs.~\cite{schaefer1,schaefer2,schaefer3,peshier,pc3} also 
forbids a Landau cut contribution since the effective quark propagators in these 
calculations do not contain any discontinuities. This stems from the fact that 
the quark self-energies in 
Refs.~\cite{schaefer1,schaefer2,schaefer3,peshier,pc3}  do not have any 
imaginary parts whereas in Refs.~\cite{kitazawa1,kitazawa2,kim} an ansatz of two 
quasiparticles was employed for spectral function based on the LQCD quark 
propagator analyzed in quenched approximation. The spectral function obtained 
with the Gribov action (\ref{gspect}) also possesses only pole contributions but 
no Landau cut.  As a result, this approach completely removes the 
quasigluons responsible for the Landau cut that should be present in a high-temperature 
quark-gluon plasma.  This is similar to findings in other nonperturbative approaches~\cite{kitazawa1,kitazawa2,kim,schaefer1,schaefer2, 
schaefer3,peshier}. We will return to the consequences
of the absence of Landau cut in the results and conclusions sections.

Returning to the problem at hand, the spectral density in (\ref{gspect}) at 
vanishing three momentum ($p \equiv |\vec{p}| =0$) contains three delta function 
singularities corresponding to the two massive modes and one new massless Gribov 
mode.  To proceed, one needs the vertex functions in presence of the Gribov 
term.  These can be determined by explicitly computing the hard-loop limit of 
the vertex function using the Gribov propagator.  One can verify, after the 
fact, that the resulting effective quark-gluon vertex function satisfies the 
necessary Slavnov-Taylor (ST) identity

\bea
(P_1-P_2)_\mu\Gamma^\mu(P_1,P_2)= S^{-1}(P_1)-S^{-1}(P_2) \ .
\label{ward_id}
\eea
The temporal and spatial parts of the modified effective quark-gluon vertex can be written as
\bea
\Gamma^0 &=& a_G ~\gamma^0 +b_G~\bf{\gamma} \cdot \hat{\bf{p}},\nn\\
\Gamma^i &=& c_G ~\gamma^i +b_G~\hat{p}^i\gamma_0+d_G~\hat{p}^i\left(\bf{\gamma} \cdot \hat{\bf{p}}\right), \label{vertex}
\eea
where the coefficients are given by
\bea
a_G &=& 1-\frac{2g^2C_F}{(2\pi)^2}\sum_\pm\int dk \, k \, \tilde{n}_\pm(k,\gamma_G)\frac{1}{\omega_1-\omega_2}
\left[\delta Q_{01}^\pm+\delta Q_{02}^\pm\right],\nn\\
b_G &=& -\frac{2g^2C_F}{(2\pi)^2}\sum_\pm\int dk \, k \, \tilde{n}_\pm(k,\gamma_G)\frac{1}{\omega_1-\omega_2}
\left[\delta Q_{11}^\pm+\delta Q_{12}^\pm\right],\nn\\
c_G &=& 1+\frac{2g^2C_F}{(2\pi)^2}\sum_\pm\int dk \, k \, \tilde{n}_\pm(k,\gamma_G)\frac{1}{3(\omega_1-\omega_2)}
\left[\delta Q_{01}^\pm+\delta Q_{02}^\pm-\delta Q_{21}^\pm-\delta Q_{22}^\pm\right],\nn\\
d_G &=& \frac{2g^2C_F}{(2\pi)^2}\sum_\pm\int dk \, k \, \tilde{n}_\pm(k,\gamma_G)\frac{1}{\omega_1-\omega_2}
\left[\delta Q_{21}^\pm+\delta Q_{22}^\pm\right],\nn
\eea
with
\bea
\delta Q_{n1}^\pm &=& Q_n(\tilde{\omega}_{11}^\pm,p)- Q_n(\tilde{\omega}_{21}^\pm,p){\rm{~for~}} n=0,1,2 \, \, ,\nn\\
\omega_{m1}^\pm &=& E_\pm^0(\omega_m+k-E_\pm^0)/k {\rm{~for~}} m=1,2\, \, , \nn\\
\omega_{m2}^\pm &=& E_\pm^0(\omega_m-k+E_\pm^0)/k {\rm{~for~}} m=1,2\,\, .\nn
\eea
Similarly, the four-point function can be obtained by computing the necessary 
diagrams in the hard-loop limit and it satisfies the following generalized ST 
identity 
\bea
P_\mu\Gamma^{\mu\nu}(-P_1,P_1;-P_2,P_2) = \Gamma^{\nu}(P_1-P_2,-P_1;P_2)-\Gamma^{\nu}(-P_1-P_2,P_1;P_2) \ . \label{wi_4pt}
\eea

\section{One-loop dilepton production with the Gribov action}
\label{dilepton}

The dilepton production rate for a dilepton with energy $\omega$ and 
three-momentum ${\vec q}$ is related to the discontinuity of the photon self 
energy $\Pi^{\mu\nu}(Q)$ as~\cite{larry} 
\bea
\frac{dR}{d\omega d^3q}=\frac{\alpha}{12\pi^3Q^2}\frac{1}{e^{\beta\omega}-1}\frac{1}{2\pi i} \, \textmd{Disc} \, \Pi_\mu^\mu(Q).
\label{dilep_def}
\eea
At one-loop order, the dilepton production rate is related to the two diagrams shown in Fig.~\ref{feyn_diag}, which can be written as 
\bea
\Pi_\mu^\mu(Q)&=&\frac{5}{3}e^2\sum_{p_0}\int\frac{d^3p}{(2\pi)^3} \biggl\{\textmd{Tr}\biggl[S(P)
~\Gamma_\mu(K,Q,-P)~S(K)~\Gamma_\mu(-K,-Q,P)\biggr]\label{se_tr} \nn \\
&& \hspace{5cm} + \; \textmd{Tr}\biggl[S(P)~\Gamma_\mu^\mu(-P,P;-Q,Q)\biggr]\biggr\}, \label{dilep_tr}
\eea
where $K=P-Q$.  The second term in (\ref{dilep_tr}) is due to the tadpole 
diagram shown in Fig.~\ref{feyn_diag} which, in the end, does not contribute 
since $\Gamma_\mu^\mu=0$. However, the tadpole diagram is essential to satisfy 
the transversality condition, $Q_\mu\Pi^{\mu\nu}(Q)=0$ and thus gauge invariance
and charge conservation in the system.

\begin{figure}[t]
\begin{center}
\includegraphics[width=0.35\linewidth]{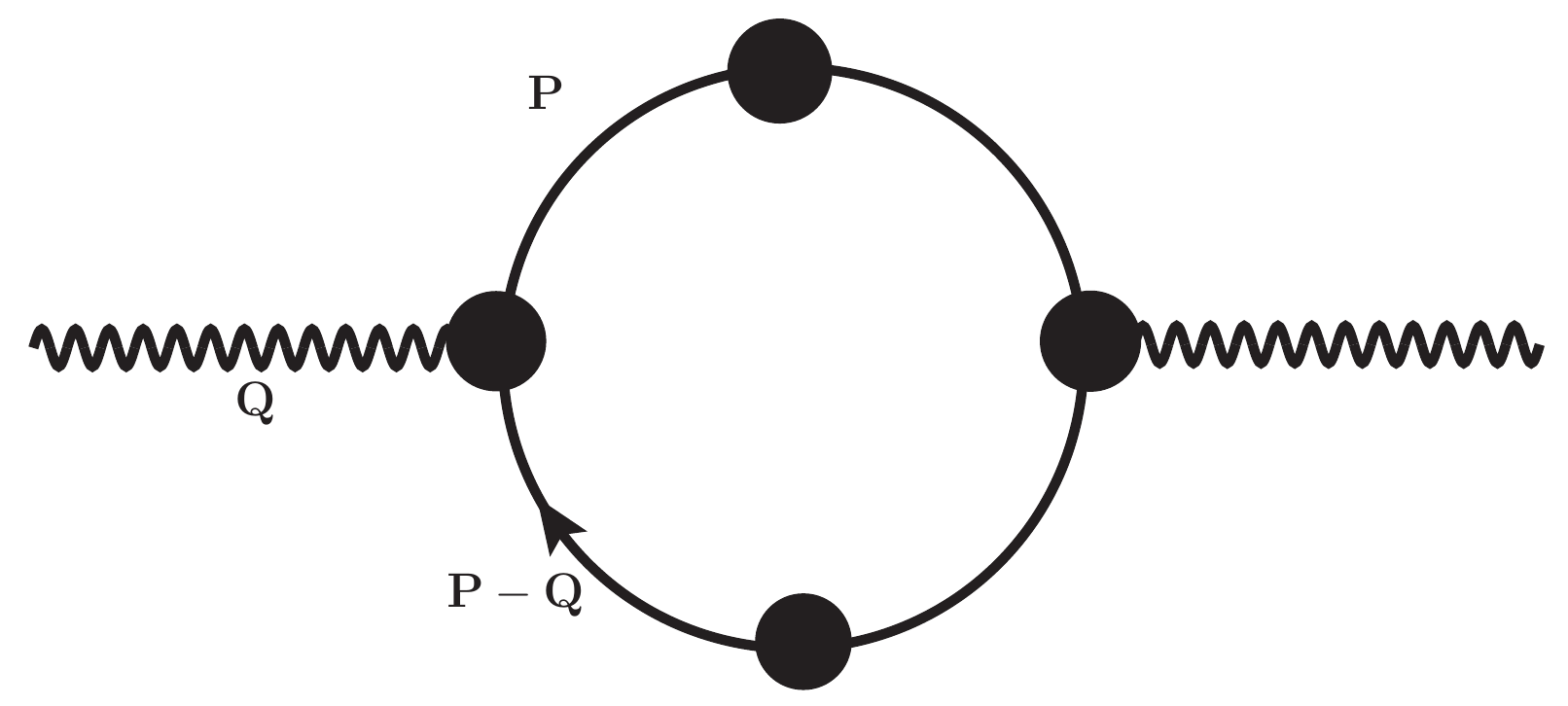}  
\hspace{5mm}
\includegraphics[width=0.3\linewidth]{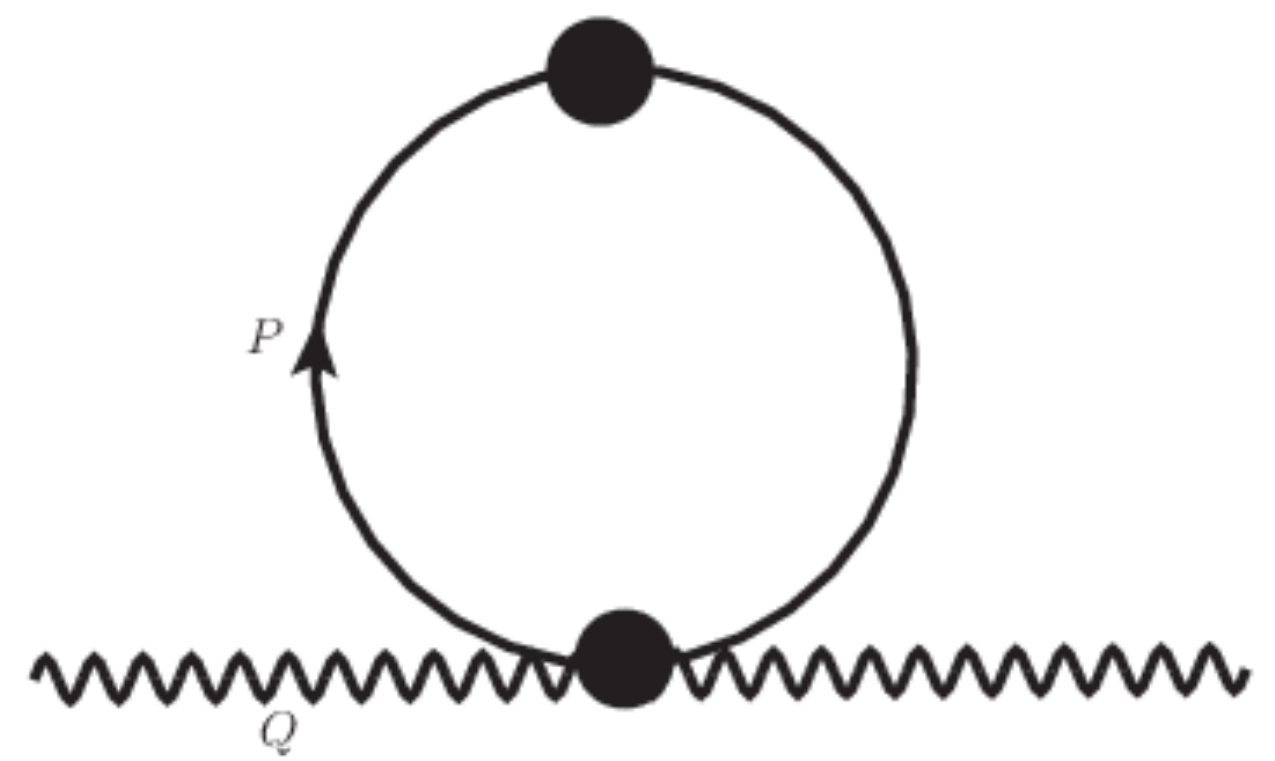}
\end{center}
\caption{The self-energy (left) and  tadpole (right) diagrams in one loop order.}
\label{feyn_diag}
\end{figure}

Using the $n$-point functions computed in sec.~\ref{setup} and performing traces, one obtains 
\bea
\Pi_\mu^\mu(\vec{q}=0)&=&\frac{10}{3}e^2T\sum_{p_0}\int\frac{d^3p}{(2\pi)^3}  \nn \\
&& \times \Biggl[\left\{\frac{(a_G+b_G)^2}{D_+(\omega_1,p,\gamma_G)D_-(\omega_2,p,\gamma_G)}+
\frac{(a_G-b_G)^2}{D_-(\omega_1,p,\gamma_G)D_+(\omega_2,p,\gamma_G)}\right\}\nn\\
&& - \left\{\frac{(c_G+b_G+d_G)^2}{D_+(\omega_1,p,\gamma_G)D_-(\omega_2,p,\gamma_G)}+
\frac{(c_G-b_G+d_G)^2}{D_-(\omega_1,p,\gamma_G)D_+(\omega_2,p,\gamma_G)}\right\}\nn\\
&& -2c_G^2\left\{\frac{1}{D_+(\omega_1,p,\gamma_G)D_+(\omega_2,p,\gamma_G)}+\frac{1}{D_-(\omega_1,p,\gamma_G)D_-(\omega_2,p,\gamma_G)}\right\}\Biggr].
\label{dlep_trace_se}
\eea

The discontinuity can be obtained by Braaten-Pisarski-Yuan (BPY) prescription~\cite{yuan}
\bea
\textmd{Disc~}T\sum_{p_0}f_1(p_0)f_2(q_0-p_0)&=& 2\pi i (1-e^{\beta \omega})\int d\omega_1 \int d\omega_2~n_F(\omega_1)
n_F(\omega_2)\nn \\
&& \times \ \delta(\omega-\omega_1-\omega_2) \ \rho_1(\omega_1)\rho_2(\omega_2), \label{bpy_pres}
\eea
which, after some work, allows one to determine the dilepton rate at zero three momentum
\bea
\frac{dR}{d\omega d^3q}({\vec q}=0)&=&\frac{10\alpha^2}{9\pi^4}\frac{1}{\omega^2}\int\limits_0^\infty p^2dp
\int\limits_{-\infty}^\infty d\omega_1 \int\limits_{-\infty}^\infty d\omega_2 n_F(\omega_1) n_F(\omega_2) 
\delta(\omega-\omega_1-\omega_2)\nn\\
&&\Bigg[4\left(1-\frac{\omega_1^2-\omega_2^2}{2p\,\omega}\right)^2 \rho_+^G(\omega_1,p) \rho_-^G(\omega_2,p)\nn\\
&&+\left(1+\frac{\omega_1^2+\omega_2^2-2p^2-2m_q^2(\gamma_G)}{2p\,\omega}\right)^2\rho_+^G(\omega_1,p) \rho_+^G(\omega_2,p)\nn\\
&&+\left(1-\frac{\omega_1^2+\omega_2^2-2p^2-2m_q^2(\gamma_G)}{2p\,\omega}\right)^2\rho_-^G(\omega_1,p) \rho_-^G(\omega_2,p)\Bigg]
.\label{dilep_spec}
\eea



Using (\ref{gspect}) and considering all physically allowed processes by the in-medium dispersion, 
the total contribution can be expressed as
\bea
\frac{dR}{d\omega d^3q}\Big\vert^{pp}({\vec q}=0)&=&\frac{10\alpha^2}{9\pi^4}\frac{1}{\omega^2}
\int\limits_0^\infty p^2\, dp \times \nn \\
&& \Biggl [\delta(\omega-2\omega_+)\ n_F^2(\omega_+)\left(\frac{\omega_+^2-p^2}{2m_q^2(\gamma_G)}\right)^2
\left\{1+\frac{\omega_+^2-p^2-m_q^2(\gamma_G)}{p~\omega}\right\}^2\nn\\
&&+~\delta(\omega-2\omega_-)\ n_F^2(\omega_-)\left(\frac{\omega_-^2-p^2}{2m_q^2(\gamma_G)}\right)^2
\left\{1-\frac{\omega_-^2-p^2-m_q^2(\gamma_G)}{p~\omega}\right\}^2\nn\\
&&+~\delta(\omega-2\omega_G)\ n_F^2(\omega_G)\left(\frac{\omega_G^2-p^2}{2m_q^2(\gamma_G)}\right)^2
\left\{1-\frac{\omega_G^2-p^2-m_q^2(\gamma_G)}{p~\omega}\right\}^2 \nn \\
&& +4 \ \delta(\omega-\omega_+-\omega_-) \ n_F(\omega_+) \ n_F(\omega_-) 
\left(\frac{\omega_+^2-p^2}{2m_q^2(\gamma_G)}\right)
\left(\frac{\omega_-^2-p^2}{2m_q^2(\gamma_G)}\right) \nn \\
&& \times \left\{1-\frac{\omega_+^2-\omega_-^2}{2p\,\omega}\right\}^2 \nn \\
&& +\delta(\omega-\omega_++\omega_-) \ n_F(\omega_+)n_F(-\omega_-)
\left(\frac{\omega_+^2-p^2}{2m_q^2(\gamma_G)}\right)\left(\frac{\omega_-^2-p^2}{2m_q^2(\gamma_G)}\right) \nn \\
&& \times \left\{1+\frac{\omega_+^2+\omega_-^2-2p^2-2m_q^2(\gamma_G)}{2p\,\omega}\right\}^2 
~\Biggr]. \label{dilep_pp}
\eea
Inspecting the arguments of the various energy conserving $\delta$-functions in 
(\ref{dilep_pp}) one can understand the physical processes originating from the 
poles of the propagator.   The first three terms in (\ref{dilep_pp}) correspond 
to the annihilation processes of $q_+{\bar q}_+\rightarrow \gamma^*$, $q_-{\bar 
q}_-\rightarrow \gamma^*$, and $q_G{\bar q}_G\rightarrow \gamma^*$, 
respectively.  The fourth term corresponds to the annihilation of $q_+{\bar 
q}_-\rightarrow \gamma^*$.  On the other hand, the fifth term corresponds to a 
process, $q_+\rightarrow q_-\gamma*$, where a $q_+$ mode makes a transition to a 
$q_-$ mode along with a virtual photon.  These processes involve soft quark 
modes ($q_+, \, q_-$, and $q_G$ and their antiparticles) which originate by 
cutting the self-energy diagram in Fig.~\ref{feyn_diag} through the internal 
lines without a ``blob''.  The virtual photon, $\gamma^*$, in all these five 
processes decays to lepton pair and can be visualized from the dispersion plot 
as displayed in the Fig.~\ref{dilepton_processes}.  The momentum integration in 
Eq.~(\ref{dilep_pp}) can be performed using the standard delta function 
identity 
\bea
\delta(f(x))&=& \sum_i \frac{\delta(x-x_i)}{\mid \! f'(x) \!\mid_{x=x_i}}, \label{delta-prop}
\eea
where $x_i$ are the solutions of $f(x_i)=0$.

\begin{figure}[t]
\begin{center}
\includegraphics[width=0.5\linewidth]{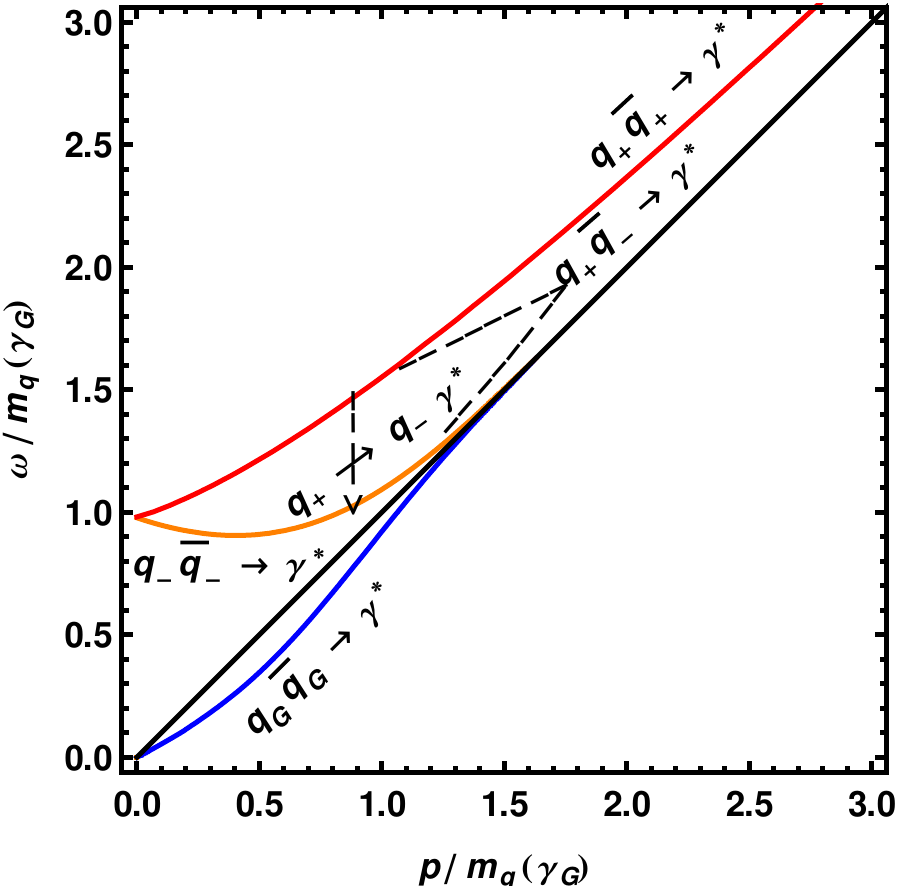}  
\end{center}
\caption{Various dilepton processes which originate from the in-medium dispersion with the Gribov term.}
\label{dilepton_processes}
\end{figure}

\begin{figure}[t]
\begin{center}
\includegraphics[width=0.75\linewidth]{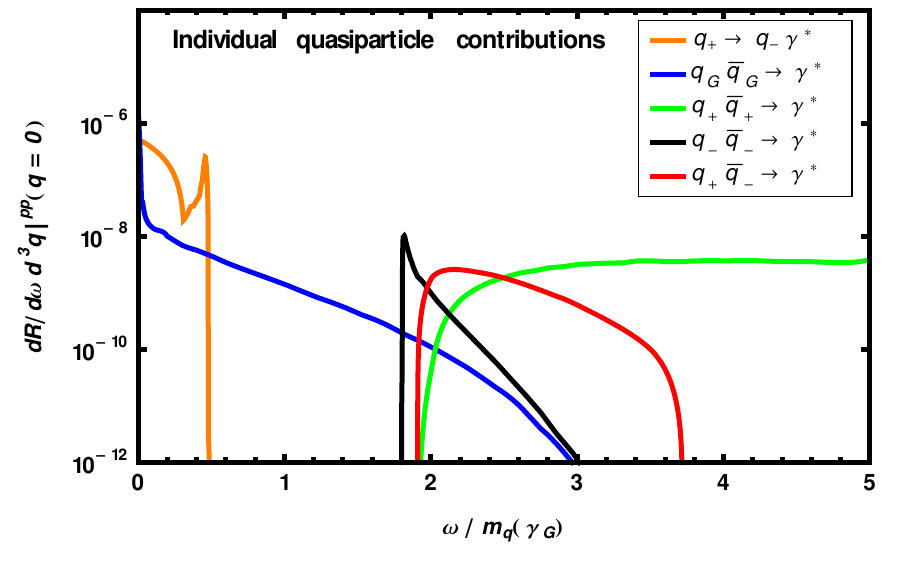} 
\end{center}
\caption{The dilepton production rates corresponding to quasiparticle processes in Fig.~\ref{dilepton_processes}.}
\label{dilepton_pp}
\end{figure}

\begin{figure}[t]
\begin{center}
\includegraphics[width=0.75\linewidth]{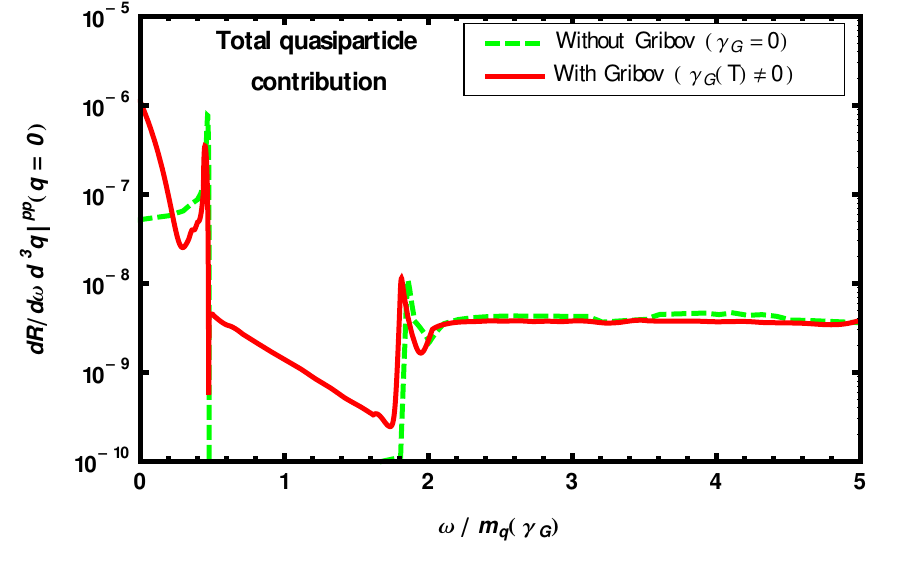} 
\end{center}
\caption{Comparison of dilepton production rates involving various quasiparticle modes with and without inclusion of $\gamma_G$.}
\label{dilepton_comparison}
\end{figure}

The contribution of various individual processes to  the dilepton production 
rate in presence of the Gribov term are displayed in the  
Fig.~\ref{dilepton_pp}.  Note that in this figure and in subsequent figures 
showing the dilepton rate, the vertical axis shows the dimensional late dilepton 
rate $dR/d^4p = dN/d^4xd^4p$ and the horizontal axis is scaled by the thermal 
quark mass as to make it dimensionless. In Fig.~\ref{dilepton_pp} we see that 
the transition process, $q_+\rightarrow q_-\gamma*$, begins at the energy 
$\omega=0$ and ends up with a van-Hove peak\,\footnote{A van-Hove 
peak~\cite{van_Hove,van_Hove1} appears where the density of states diverges as 
$f'(x)|_{x=x_0}=0$ since the density of states is inversely proportional to 
$f'(x)$.} where all of the transitions from $q_+$ branch are directed towards 
the minimum of the $q_-$ branch. The annihilation process involving the massless 
spacelike Gribov modes, $q_G{\bar q}_G\rightarrow \gamma^*$, also starts at 
$\omega=0$ and falls-off very quickly.  The annihilation of the two plasmino 
modes, $q_-{\bar q}_-\rightarrow \gamma^*$, opens up with again a van-Hove peak 
at $\omega=2 \times $ the minimum energy of the plasmino mode. The contribution 
of this process decreases exponentially.  At $\omega=2m_q(\gamma_G)$, the 
annihilation processes involving usual quark modes, $q_+{\bar q}_+\rightarrow 
\gamma^*$, and that of a quark and a plasmino mode, $q_+{\bar q}_-\rightarrow 
\gamma^*$, begin.  However, the former one ($q_+{\bar q}_+\rightarrow \gamma^*$) 
grows with the energy and would converge to the usual Born rate (leading order 
perturbative rate)~\cite{born} at high mass whereas the later one ($q_+{\bar 
q}_-\rightarrow \gamma^*$) initially grows at a very fast rate, but then 
decreases slowly and finally drops very quickly. The behavior of the latter 
process can easily be understood from the dispersion properties of quark and 
plasmino mode. Summing up, the total contribution of all theses five processes 
is displayed in Fig.~\ref{dilepton_comparison}. This is compared with the 
similar dispersive contribution
when $\gamma_G=0$~\cite{yuan}, comprising processes $q_+\rightarrow q_-\gamma*$, 
$q_+{\bar q}_+\rightarrow \gamma^*$, $q_-{\bar q}_-\rightarrow \gamma^*$ and  
$q_+{\bar q}_-\rightarrow \gamma^*$. We note that when $\gamma_G=0$, the 
dilepton rate contains both van-Hove peaks and an energy gap~\cite{yuan}. In 
presence of the Gribov term ($\gamma_G\ne 0$), 
the van-Hove peaks remain, but the energy gap disappears due to the annihilation 
of new massless Gribov modes, $q_G{\bar q}_G\rightarrow \gamma^*$.

\begin{figure}[t]
\begin{center}
\includegraphics[width=0.75\linewidth]{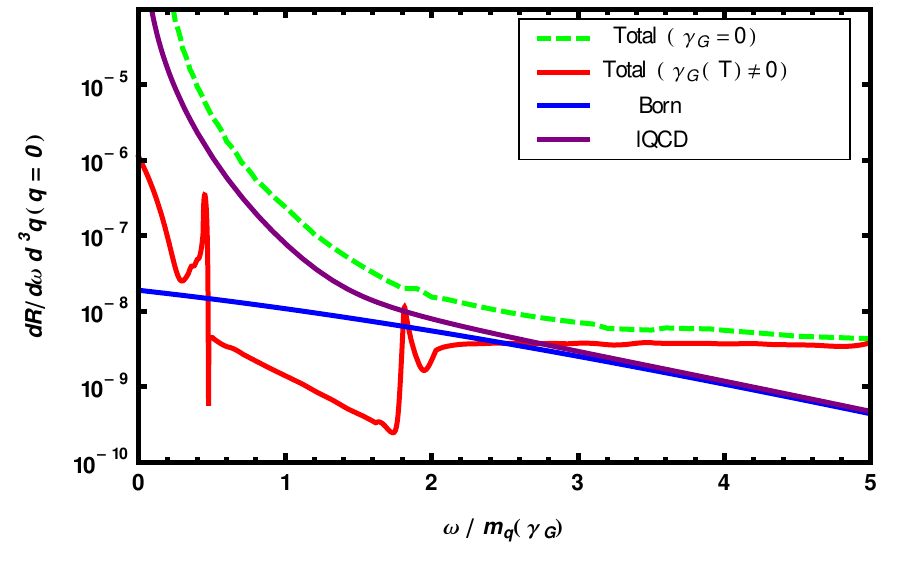} 
\end{center}
\caption{Comparison of various dilepton production  rates from  the deconfined matter.}
\label{dilepton_all}
\end{figure}

In Fig.~\ref{dilepton_all} we compare the rates obtained using various 
approximations:  leading-order perturbative (Born) rate~\cite{born}, quenched 
LQCD rate~\cite{ding,laermann}, and with and without the Gribov term.  The 
non-perturbative rate with the Gribov term shows important structures compared 
to the Born rate at low energies.  But when compared to the total HTLpt 
rate~\footnote{Since HTL spectral function (i.e, $\gamma_G=0$) has both pole and 
Landau cut contribution, so the HTLpt rate~\cite{yuan} contains an additional 
{\textit {higher order}} contribution due to the Landau cut stemming from 
spacelike momenta.} it is suppressed  in the low mass region due to the absence 
of Landau cut contribution for $\gamma_G \neq 0$.  It seems as if the higher 
order Landau cut contribution due to spacelike momenta for $\gamma_G=0$ is 
replaced by the soft process involving spacelike Gribov modes in the collective 
excitations for $\gamma_G\neq 0$. We also note that the dilepton rate~\cite{kim} 
using the spectral function constructed with two pole ansatz by analyzing LQCD 
propagator in quenched approximation~\cite{kitazawa1,kitazawa2} shows similar 
structure  as found here for $\gamma_G\ne0$. On the other hand, such structure 
at low mass is also expected in the direct computation of dilepton rate from 
LQCD in quenched approximation~\cite{ding,laermann}. However, a smooth variation 
of the rate was found at low mass. The computation of dilepton rate in LQCD 
involves  
various intricacies and uncertainties.  This is because, as noted in 
sec.~\ref{intro}, the spectral function in continuous time is obtained from the 
correlator in finite set of discrete Euclidean time using a probabilistic MEM 
method~\cite{mem1,mem2,mem3} with a somewhat ad hoc continuous ansatz for the 
spectral function at low energy and also fundamental difficulties in performing 
the necessary analytic continuation in LQCD. Until LQCD overcomes the 
uncertainties and difficulties in the computation of the vector spectral 
function, one needs to depend, at this juncture, on the prediction of the 
effective approaches for dilepton rate at low mass in particular. We further 
note that at high-energies the rate for both  $\gamma_G=0$ and $\gamma_G\ne 0$ 
is higher than the lattice data and Born rate.  This is a consequence of using 
the HTL self-energy also at high-energies/momentum where the soft-scale 
approximation breaks down. Nevertheless, the low mass rate obtained here by 
employing the non-perturbative magnetic scale ($\gamma_G\ne 0$) in addition to 
the electric scale allows for a model-based inclusion of the effect of 
confinement and the result has a somewhat rich structure at low energy compared 
to that obtained using only the electric scale ($\gamma_G = 0$) as well in 
LQCD. 

We make some general comments concerning the dilepton rate below.
 If one looks at the dispersion plots in Fig.~\ref{disp_rel} for  $\gamma_G=0$, 
then one finds that $\omega_-$  falls off  exponentially  and approaches light 
cone, whereas $\omega_+$ does not follow fall off exponentially to light cone, but 
instead behaves as $[p+m^2_q(T)/p]$ for large $p$. On the other hand,  in 
the presence of  $\gamma_G \ne 0$ both  $\omega_-$  and $\omega_G$ approach the light 
cone very quickly, but again $\omega_+$ has a similar asymptotic behavior as 
before.
This feature of $\omega_+$  makes the dilepton rate at large $\omega$ in 
Fig.~\ref{dilepton_all} saturate for both $\gamma_G=0$ and $\gamma_G \ne 0$, 
because the dominant contribution comes from the annihilation of two $\omega_+$
as discussed in Fig.~\ref{dilepton_pp}. In general, the total dilepton rate in 
Fig.~\ref{dilepton_all}, behaves as  $\sim \exp(-\omega/T)$ for  
$\gamma_G(T) = 0$  due to the Landau damping contribution coming from 
the quasigluons in a hot and dense medium. As the Landau cut contribution is 
missing in the $\gamma_G(T) \neq 0$ case, one finds a leveling off at low $\omega$.
In other words, since the Landau damping contribution is absent for 
$\gamma_G(T) \ne 0$, the rate approaches that of the pole-pole 
contribution for $\gamma_G=0$ as shown in Fig.~\ref{dilepton_comparison},  
except in the mass gap region.  We further note that the LQCD 
rate~\cite{ding} matches with Born rate at large $\omega$  simply because a free spectral function 
has been assumed for large $\omega$. On the other hand the LQCD 
spectral function~\cite{ding} at low $\omega$ is sensitive to the prior 
assumptions and, in such a case, the spectral function extracted using  
a MEM~\cite{mem1,mem2,mem3} analyses should be interpreted carefully with a proper 
error analysis~\cite{mem1}. Since the MEM analyses is sensitive to the prior 
assumption, but is not very sensitive to the structure of the spectral function 
at small $\omega$, the error is expected to be significant at small 
$\omega$. The existence of fine structure such as van Hove singularities at 
small $\omega$ cannot be excluded based on the LQCD rate~\cite{ding} at this moment in time.

\section{One-loop quark number susceptibility with the Gribov action}
\label{qns}

We now turn to the computation of the quark number susceptibility (QNS) 
including the Gribov term.  The QNS can be interpreted as the response of the 
conserved quark number density, $n$ with infinitesimal variation in the quark 
chemical potentials $\mu+\delta\mu$. In QCD thermodynamics it is defined as the 
second order derivative of pressure ${\cal P}$ with respect to quark chemical 
potential, $\mu$. But again, using the fluctuation-dissipation (FD) theorem, the 
QNS for a given quark flavor can also be defined from the time-time component of 
the current-current correlator in the
vector channel~\cite{purnendu1,purnendu3,forster,kunihiro}. The QNS is in general expressed as
\bea
\chi_q(T)&=& \! \! \left. \frac{\partial n}{\partial\mu}\right|_{\mu\rightarrow 0}
\!\!\! = \!\!\!\left. \frac{\partial^2 {\cal P}}{\partial^2\mu}\right|_{\mu\rightarrow 0}
\!\!\! =\int d^4x\langle J_0(0,{\vec x})J_0(0,{\vec 0})\rangle \nn \\
&=&
\beta\int\limits_{-\infty}^{\infty}\frac{d\omega}{2\pi}\frac{-2}{1-e^{-\beta\omega}}
~\textmd{Im}~\Pi_{00}(\omega, {\vec 0}), \label{def_qns}
\eea
where $J_0$ is the temporal component of the vector current and $\Pi_{00}$ is the time-time component of the vector 
correlator or self-energy with external four-momenta $Q \equiv (\omega, \vec q)$. The above relation in
(\ref{def_qns}) is known as the  thermodynamic sum rule~\cite{forster,kunihiro} where the thermodynamic derivative 
with respect to the external source, $\mu$ is related to the time-time 
component of static correlation function in the vector channel.

In order to compute the QNS we need to calculate the imaginary part of the temporal component of the 
two one-loop diagrams given in Fig.~\ref{feyn_diag}. The contribution of the self energy diagram is
\bea
\Pi^s_{00}(Q)&=&N_fN_cT\sum_{p_0}\int\frac{d^3p}{(2\pi)^3} \textmd{Tr}\left[S(P)~\Gamma^0(K,Q,-P)~S(K)
~\Gamma^0(-K,-Q,P)\right],
\eea
where $K=P-Q$. After performing the traces of the self energy diagram, one obtains
\bea
\Pi^s_{00}(\vec{q}=0)&=& 2N_fN_cT\sum_{p_0}\int \!\!\!\frac{d^3p}{(2\pi)^3} 
\left[\frac{(a_G+b_G)^2}{D_+(\omega_1,p,\gamma_G)D_-(\omega_2,p,\gamma_G)}\right.\nonumber\\
&&\hspace{3.3cm}+\left.\frac{(a_G-b_G)^2}{D_-(\omega_1,p,\gamma_G)D_+(\omega_2,p,\gamma_G)}\right] , \label{se}
\eea
where
\bea
a_G + b_G &=& 1-\frac{2g^2C_F}{(2\pi)^2}\sum_\pm\int dk k \tilde{n}_\pm(k,\gamma_G)\frac{1}{\omega}\nn\\
&& \times \Bigl[Q_0(\tilde{\omega}_{11}^\pm,p) + Q_1(\tilde{\omega}_{11}^\pm,p)+Q_0(\tilde{\omega}_{21}^\pm,p)
- Q_1(\tilde{\omega}_{21}^\pm,p)\nn\\
&&+Q_0(\tilde{\omega}_{12}^\pm,p) + Q_1(\tilde{\omega}_{12}^\pm,p)+Q_0(\tilde{\omega}_{22}^\pm,p)- Q_1(\tilde{\omega}_{22}^\pm,p)\Bigr]\nn\\
&=& 1+\frac{1}{\omega}\left[D_+(\omega_1,p,\gamma_G)+D_-(\omega_2,p,\gamma_G)-\omega_1-\omega_2\right]\nn\\
&=& 1- \frac{\omega_1+\omega_2}{\omega} + \frac{D_+(\omega_1,p,\gamma_G)+D_-(\omega_2,p,\gamma_G)}{\omega}, \label{coeffsa}
\eea
and
\bea
a_G - b_G &=& 1-\frac{2g^2C_F}{(2\pi)^2}\sum_\pm\int dk k \tilde{n}_\pm(k,\gamma_G)\frac{1}{\omega}\nn\\
&& \times \Bigl[Q_0(\tilde{\omega}_{11}^\pm,p)- Q_1(\tilde{\omega}_{11}^\pm,p)+Q_0(\tilde{\omega}_{21}^\pm,p)
+ Q_1(\tilde{\omega}_{21}^\pm,p)\nn\\
&&+Q_0(\tilde{\omega}_{12}^\pm,p) - Q_1(\tilde{\omega}_{12}^\pm,p)+Q_0(\tilde{\omega}_{22}^\pm,p) + Q_1(\tilde{\omega}_{22}^\pm,p)\Bigr]\nn\\
&=& 1+\frac{1}{\omega}\left[D_-(\omega_1,p,\gamma_G)+D_+(\omega_2,p,\gamma_G)-\omega_1-\omega_2\right]\nn\\
&=& 1- \frac{\omega_1+\omega_2}{\omega} + \frac{D_-(\omega_1,p,\gamma_G)+D_+(\omega_2,p,\gamma_G)}{\omega}, \label{coeffsb}
\eea
where $D_\mp(\omega,p,\gamma_G)$ were defined in Eq.~(\ref{dpm}). We write only those terms of Eq.~(\ref{se}) which contain discontinuities
\bea
\frac{(a_G+b_G)^2}{D_+(\omega_1,p,\gamma_G)D_-(\omega_2,p,\gamma_G)} &=& \frac{(1- \frac{\omega_1+\omega_2}{\omega})^2}
{D_+(\omega_1,p,\gamma_G)D_-(\omega_2,p,\gamma_G)}\nonumber\\
&& \hspace{5mm} + \frac{1}{\omega^2}\left\{\frac{D_+(\omega_1,p,\gamma_G)}{D_-(\omega_2,p,\gamma_G)}+\frac{D_-(\omega_2,p,\gamma_G)}
{D_+(\omega_1,p,\gamma_G)}\right\}, \nn\\
\frac{(a_G-b_G)^2}{D_-(\omega_1,p,\gamma_G)D_+(\omega_2,p,\gamma_G)} &=& \frac{(1- \frac{\omega_1+\omega_2}{\omega})^2}
{D_-(\omega_1,p,\gamma_G)D_+(\omega_2,p,\gamma_G)}\nonumber\\
&& \hspace{5mm} + \frac{1}{\omega^2}\left\{\frac{D_-(\omega_1,p,\gamma_G)}{D_+(\omega_2,p,\gamma_G)}+\frac{D_+(\omega_2,p,\gamma_G)}
{D_-(\omega_1,p,\gamma_G)}\right\}.\label{coeffs1}
\eea
Calculating the discontinuity using the BPY prescription given in Eq.~(\ref{bpy_pres}), one can write the imaginary part of Eq.~(\ref{se}) as
\bea
\textmd{Im}~\Pi^s_{00} &=& 4N_cN_f\pi(1-e^{\beta\omega})\int\frac{d^3p}{(2\pi)^3}\int d\omega_1 
\int d\omega_2 ~\delta(\omega-\omega_1-\omega_2)
n_F(\omega_1)n_F(\omega_2)\nn\\
&& \times \Bigl[\left(1- \frac{\omega_1+\omega_2}{\omega}\right)^2 \rho_+^G(\omega_1,p)\rho_-^G(\omega_2,p)+
\frac{C_1\rho_+^G(\omega_2,p)+C_2\rho_-^G(\omega_2,p)}{\omega^2}\Bigr], \label{im_se}
\eea
with
\bea
C_1 &=& \textmd{Im}~D_-(\omega_1,p)=0, \nn\\
C_2 &=& \textmd{Im}~D_+(\omega_1,p)=0. \label{c1c2}
\eea
The tadpole part of Fig.~\ref{feyn_diag} can now be written as
\bea
\Pi^t_{00}(Q)&=&N_fN_cT\sum_{p_0}\int\frac{d^3p}{(2\pi)^3} 
 \textmd{Tr}\biggl[S(P)~\Gamma_{00}(-P,P;-Q,Q)\biggr]. \label{qns_tad}
\eea
The four-point function $\Gamma_{00}$ at zero three-momentum can be obtained using Eq.~(\ref{wi_4pt}) giving
\bea
\Gamma^{00}&=& -(e_G\gamma^0+f_G~ \hat{p}\cdot\vec{\gamma}),\label{4pt}\\
e_G &=& \frac{2g^2c_F}{(2\pi)^2}\sum_\pm\int dk k \tilde{n}_\pm(k,\gamma_G)\frac{1}{(\omega_1-\omega_2)}\left[\delta Q_{01}^\pm+\delta Q_{02}^\pm+\delta Q_{01}^{\pm\prime}+\delta Q_{02}^{\pm\prime}\right],\nn\\
f_G &=& \frac{2g^2c_F}{(2\pi)^2}\sum_\pm\int dk k \tilde{n}_\pm(k,\gamma_G)\frac{1}{(\omega_1-\omega_2)}\left[\delta Q_{11}^\pm+\delta Q_{12}^\pm+\delta Q_{11}^{\pm\prime}+\delta Q_{12}^{\pm\prime}\right],\nn
\eea
where
\bea
\delta Q_{n1}^{\pm\prime} &=& Q_n(\tilde{\omega}_{11}^\pm,p)- Q_n(\tilde{\omega}_{21}^{\pm\prime},p){\rm{~for~}} n=0,1,2 \, \, , \nn\\
\tilde{\omega}_{21}^{\pm\prime}&=& E_\pm^0(\omega_2^\prime +k-E_\pm^0)/k  \, , \nn\\
\tilde{\omega}_{22}^{\pm\prime}&=& E_\pm^0(\omega_2^\prime -k+E_\pm^0)/k  \, , \nn\\
\omega_2^\prime &=& \omega_1 + \omega. \nn
\eea

Proceeding in a similar way as the self-energy diagram, the contribution from the tadpole diagram is
\bea
\textmd{Im}~\Pi_{00}^t &=& -4N_cN_f\pi(1-e^{\beta\omega})\int\frac{d^3p}{(2\pi)^3}\int d\omega_1 \int d\omega_2
~ \delta(\omega-\omega_1-\omega_2)\frac{n_F(\omega_1)n_F(\omega_2)}{\omega^2}\nn\\
&&\times \Bigl[C_1\rho_+^G(\omega_2,p)+C_2\rho_-^G(\omega_2,p)\Bigr]=0. \label{tad_im}
\eea
The total imaginary contribution of the temporal part shown in Fig.~\ref{feyn_diag} can now be written as 
\bea
\textmd{Im}~\Pi_{00} &=&  \textmd{Im}~\Pi^s_{00} + \textmd{Im}~\Pi^t_{00} \nn \\
&=& 4N_cN_f\pi(1-e^{\beta\omega})\int\frac{d^3p}{(2\pi)^3}\int d\omega_1 \int d\omega_2 ~\delta(\omega-\omega_1-\omega_2)
n_F(\omega_1)n_F(\omega_2)\nn\\
&& \times \Bigl[\left(1- \frac{\omega_1+\omega_2}{\omega}\right)^2 \rho_+^G(\omega_1,p)\rho_-^G(\omega_2,p)\Bigr]. \label{tot_im}
\eea
It is clear from (\ref{im_se}) and (\ref{tad_im}) that the tadpole contribution 
in (\ref{tad_im}) exactly cancels with the second term of (\ref{im_se}) even if 
$C_1$ and $C_2$ are finite, e.g., for the HTL
case ($\gamma_G=0$) \cite{purnendu1,purnendu3}. At finite $\gamma_G$, the form 
of the sum of self-energy and tadpole diagrams remains the same, even though the 
individual contribution are modified.

Putting this in the expression for the QNS in Eq.~(\ref{def_qns}), we obtain
\bea
\chi_q(T)&=&4N_cN_f\beta\int\frac{d^3p}{(2\pi)^3}\int\limits_{-\infty}^{\infty}d\omega\int d\omega_1 
\int d\omega_2 ~\delta(\omega-\omega_1-\omega_2)n_F(\omega_1)n_F(\omega_2) \nn \\
&&\hspace{1cm} \times \Big[\left(1- \frac{\omega_1+\omega_2}{\omega}\right)^2 
\rho_+^G(\omega_1,p)\rho_-^G(\omega_2,p)\Big] \nn \\
%
&=& 4N_cN_f\beta\int\frac{d^3p}{(2\pi)^3}\Bigl[
\left(\frac{\omega_+^2-p^2}{2m_q^2(\gamma_G)}\right)^2n_F(\omega_+)n_F(-\omega_+) \nn\\
&+&\left(\frac{\omega_-^2-p^2}{2m_q^2(\gamma_G)}\right)^2n_F(\omega_-)n_F(-\omega_-)+\left(\frac{\omega_G^2-p^2}{2m_q^2(\gamma_G)}\right)^2n_F(\omega_G)n_F(-\omega_G)
\Bigr] \nn \\
&=& \chi_q^{\rm{pp}}(T).\label{qns_pp}
\eea
where we represent the total $\chi_q(T)$ as $\chi^{\rm {pp}}_q(T)$ since there 
is only the pole-pole contribution for $\gamma_G\ne 0$. However for $\gamma_G=0$
there will be pole-cut ($\chi^{\rm {pc}}_q(T)$) and cut-cut 
($\chi^{\rm{cc}}_q(T))$ contribution in addition to pole-pole contribution 
because the spectral
function contains  pole part + Landau cut contribution of the quark propagator.

\begin{figure}[t]
\begin{center}
\includegraphics[width=0.48\linewidth]{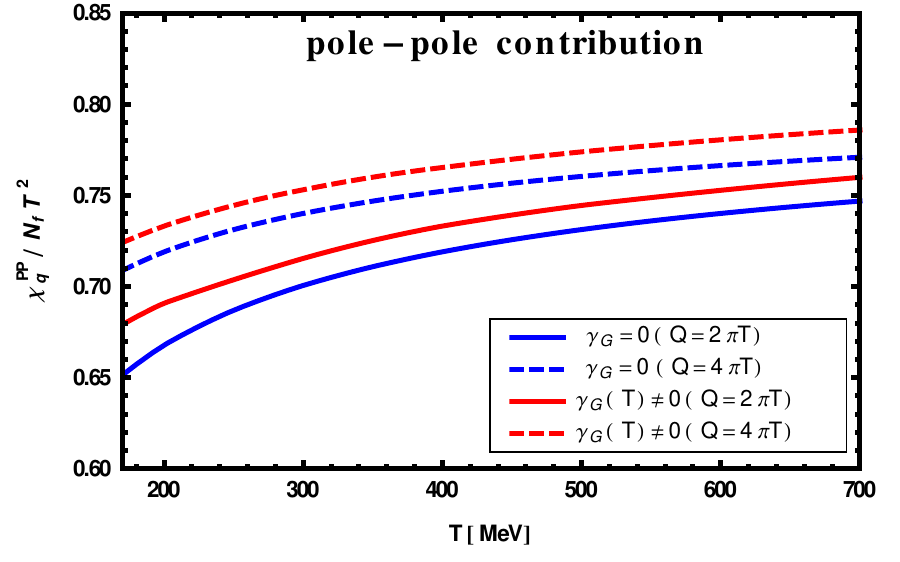}  
\hspace{2mm}
\includegraphics[width=0.48\linewidth]{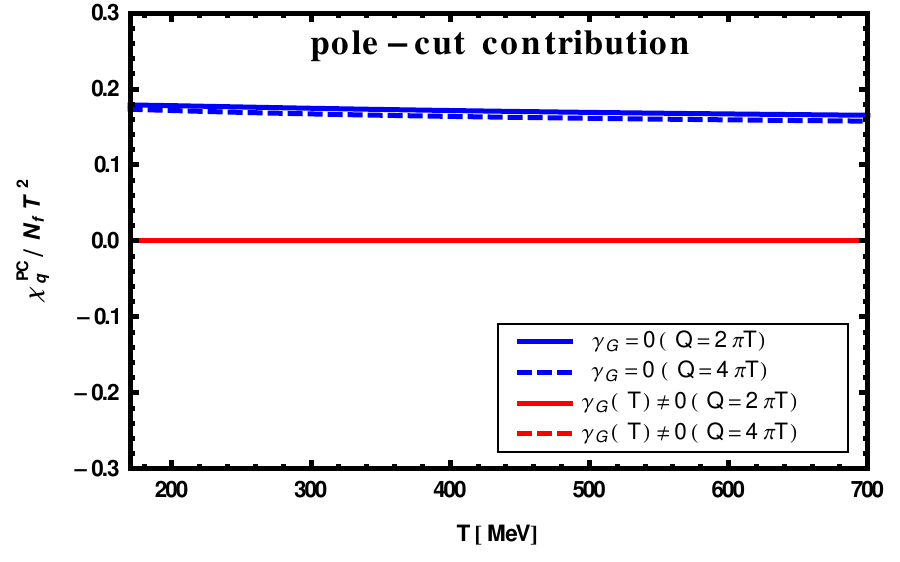}\\
\vspace{5mm}
\includegraphics[width=0.48\linewidth]{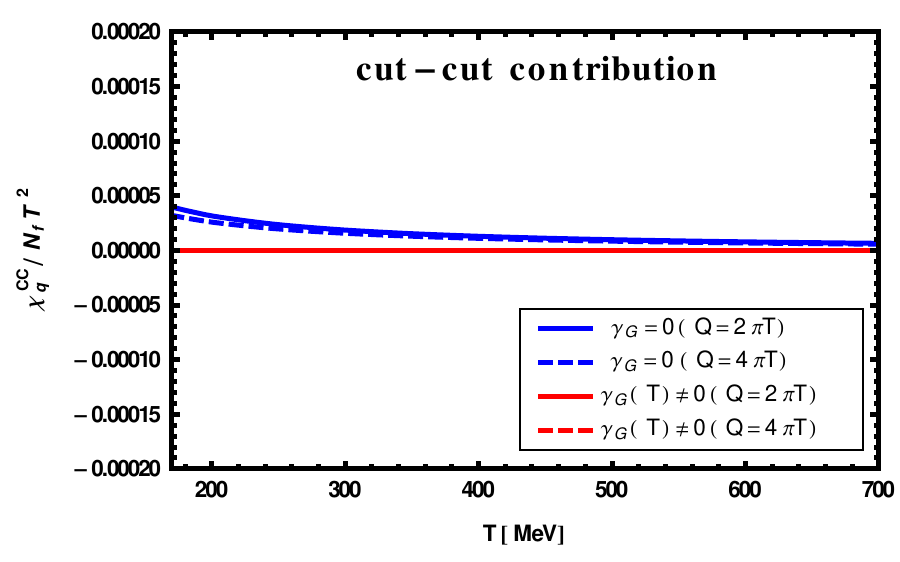}  
\hspace{2mm}
\includegraphics[width=0.48\linewidth]{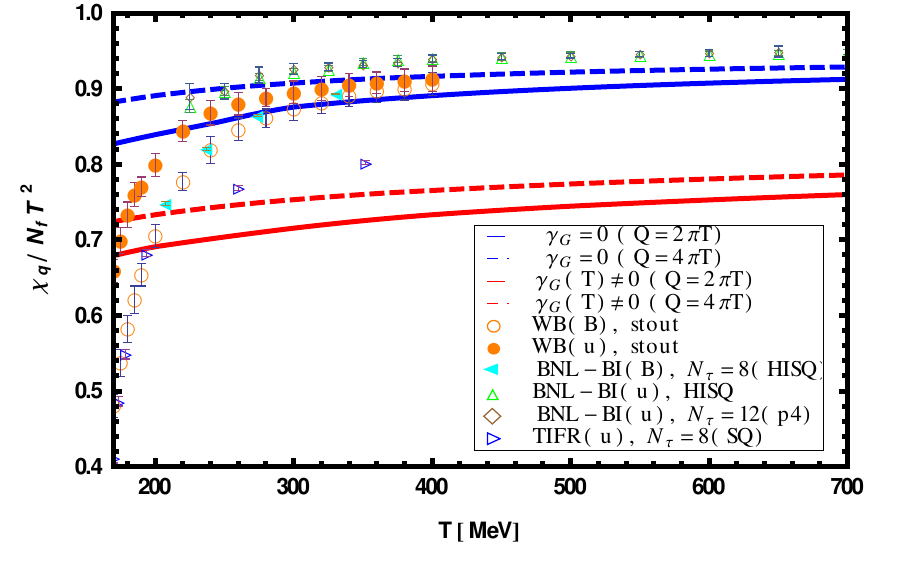}
\end{center}
\caption{ QNS scaled with free values are compared with and without the 
inclusion of $\gamma_G$. In each case a band appears due to the choice of the 
two renormalization scales as $2\pi T$ and $4\pi T $.  The various symbols 
correspond to LQCD data from various groups labeled as  WB~\cite{borsanyi2}, 
BNL-BI(B) and BNL-BI(u)~\cite{bnlb1,bnlb2}, and TIFR~\cite{tifr}.}

\label{qns_Gribov}
\end{figure}

In Fig.~\ref{qns_Gribov} we have presented the different contribution of QNS 
scaled with the corresponding free values with and without the Gribov term. 
We, at first, note that the running
coupling in (\ref{alpha_s}) is a smooth function of $T$ around and below 
$T_c$.  We have extended to low temperatures as an extrapolation of our 
high-temperature result even though our treatment is strictly not valid below $T_c$. 
Now from the first panel of Fig.~\ref{qns_Gribov}, the pole-pole contribution 
to the QNS with the Gribov action is increased at low $T$, compared to that in 
absence of the Gribov term. This improvement at low $T$ is solely due to the 
presence of the non-perturbative Gribov mode in the collective excitations. 
However, at high $T$ both contributions become almost same as the Gribov mode 
disappears.  There are no pole-cut (pc) or cut-cut (cc) contribution for 
$\gamma_G(T)\ne 0$, compared to that for $\gamma_G = 0$. The pc and cc 
contributions in absence of magnetic scale are displayed in second and third 
panels. As a result, we find that the QNS in presence of magnetic scale contains 
only the pp-contribution due to collective excitations originating from the 
in-medium dispersion whereas, in absence of magnetic scale, the QNS is enhanced 
due to additional higher order Landau cut (i.e., pole-cut + cut-cut) 
contribution as shown in the fourth panel. When compared with LQCD data from 
various groups~\cite{borsanyi2,bnlb1,bnlb2,tifr}, the QNS in presence of 
magnetic scale lies around $(10-15)\%$ below the LQCD results whereas that in 
absence of magnetic scale is very close to LQCD data. This is expected due to 
the {\textit{additional higher-order Landau cut}} contribution  in absence of 
magnetic scale as discussed earlier. This also suggests that it is necessary to 
include higher loop orders in QNS in presence of the   magnetic scale, which is  
beyond scope of this paper. However, we hope to carry out this non-trivial task 
in near future.

\section{Conclusions and outlook}
\label{conclusions}

In this paper we considered the effect of inclusion of magnetic screening in the 
context of the Gribov-Zwanziger picture of confinement.  In covariant gauge, 
this was accomplished by adding a masslike parameter, the Gribov parameter, to 
the bare gluon propagator resulting in the non-propagation of gluonic modes.  
Following Ref.~\cite{nansu1} we obtained the resummed quark propagator taking 
into account the Gribov parameter.  A new key feature of the resulting resummed 
quark propagator is that it contains no discontinuities.  In the standard 
perturbative hard-thermal loop approach there are discontinuities at spacelike 
momentum associated with Landau damping which seem to be absent in the GZ-HTL 
approach.  Using the resulting quark propagator, we evaluated the spectral 
function, finding that it only contains poles for $\gamma_G \neq 0$.  We then 
used these results to compute (1) the dilepton production rate at vanishing 
three-momentum and (2) the quark number susceptibility.  For the dilepton 
production rate, we found that, due to the absence of Landau damping for 
$\gamma_G \neq 0$, the rate contains sharp structures, e.g. Van Hove 
singularities, which don't seem to be present in the lattice data.  That being 
said, since the lattice calculations used a perturbative ansatz for the spectral 
function when performing their MEM analysis\cite{mem1} of the spectral 
function, it is unclear how changing the underlying prior assumptions about the 
spectral function would affect the final lattice results. 
Moreover, the error analsis for spectral function with MEM 
prescription~\cite{mem2} has to be done carefully than it was done in LQCD 
calculation~\cite{ding}. Since the result is sensitive to the prior
assumptions, the error seems to become large and as a result no conclusion
can be drawn for fine structures at low mass dileptons from the LQCD result. 
For the quark number susceptibilities, we found that, again due to the absence 
of Landau damping for $\gamma_G \neq 0$, the results do not agree well with 
available lattice data.  This can be contrasted with a standard HTLpt 
calculation, which seems to describe the lattice data quite well with no free 
parameters. It is possible that higher-order loop calculations could improve 
the agreement between the Gribov-scenario results and the lattice data; 
however, the 
success of HTLpt compared to lattice data as well as nonperturbative model 
calculations suggests that at \mbox{$T\gtrsim$ 200 MeV} the electric sector 
alone provides an accurate description of QGP thermodynamics.
Nevertheless, the present HTLpt results poses a serious 
challenge to the Gribov scenario for only inclusion of magnetic mass effects in 
the QGP.  
The absence of quasigluons responsible for the Landau cut makes 
the results for both dilepton production and quark number susceptibility 
dramatically different from those in perturbative approaches. We conclude that 
the results with present GZ action is in conflict with those in perturbative 
approaches due to the absence of the Landau cut contribution in the non-perturbative 
quark propagator.

\acknowledgments{
We would like to thank N. Su, K. Tywoniuk, W. Florkowski, and P. Petreczky for useful 
conversations.  A.~Bandyopadhyay and M.G.~Mustafa were supported by the Indian 
Department of Atomic Energy.  N.~Haque was supported the Indian Department of 
Atomic Energy and an award from the Kent State University Office of Research and 
Sponsored Programs.  M.~Strickland was supported by the U.S. Department of 
Energy under Award No.~DE-SC0013470. 

}

\end{document}